\documentclass[12pt]{iopart}

\usepackage{dsfont}
\usepackage{braket}
\usepackage{graphicx}
\usepackage{iopams}
\usepackage{amssymb}
\usepackage{color}
\newcommand{\an}{\textcolor{black}}
\begin{document}

\title[Quantum kernels for classifying dynamical singularities in a multiqubit system]{Quantum kernels for classifying dynamical singularities in a multiqubit system}

\author{Diego Tancara$^1$, José Fredes$^2$ and Ariel Norambuena$^{\ast,1}$}

\address{$^1$ Centro Multidisciplinario de Física, Universidad Mayor, Camino la Piramide 5750, Huechuraba, Santiago, Chile}
\address{$^2$  Escuela de Ingenier\'ia Civil en Computaci\'on e Inform\'atica, Universidad Mayor, Santiago, Chile}
\ead{ariel.norambuena@umayor.cl}
\vspace{10pt}
\begin{indented}
\item[]August 2017
\end{indented}

\begin{abstract}
Dynamical quantum phase transition is a critical phenomenon involving out-of-equilibrium states and broken symmetries without classical analogy. However, when finite-sized systems are analyzed, dynamical singularities of the rate function can appear, leading to a challenging physical characterization when parameters are changed. Here, we report a quantum support vector machine (QSVM) algorithm that uses quantum Kernels to classify dynamical singularities of the rate function for a multiqubit system. We illustrate our approach using $N$ long-range interacting qubits subjected to an arbitrary magnetic field, which induces a quench dynamics. Inspired by physical arguments, we introduce two different quantum Kernels, one inspired by the ground state manifold and the other based on a single state tomography. Our accuracy and adaptability results show that this quantum dynamical critical problem can be efficiently solved using physically inspiring quantum Kernels. Moreover, we extend our results for the case of time-dependent fields, quantum master equation, and when we increase the number of qubits.
\end{abstract}

\vspace{2pc}

\section{Introduction}
One intriguing critical phenomenon in finite quantum systems is the presence of dynamical singularities of the rate function~\cite{Heyl2018,Zeng2023}. In closed quantum systems, these dynamical singularities can be characterized by the emergence of zero points of the Loschmidt echo (LE) $\mathcal{L}(t) = |\langle \Psi(0)| \Psi(t) \rangle|^2$~\cite{Heyl2013}, where $\ket{\Psi(t)}$ is the evolved state after a quench dynamics (sudden change of a parameter). It is important to mention that if these dynamical singularities persist in the thermodynamical limit, we can adequately characterize a dynamical quantum phase transition (DQPT)~\cite{Zeng2023,Heyl2013,Jurcevic2017,Zhou2021}. Theoretically, the existence of dynamical singularities of the rate function in a finite system is a necessary (not sufficient) condition for having DQPT, as explained in Ref.~\cite{Zeng2023}. In general, to analyze the existence of dynamical singularities of the rate function in a system with $N$ degrees of freedom (with finite $N$), one introduces the rate function $\lambda(t) = -(1/N)\mbox{log}[\mathcal{L}(t)]$, where the initial state $\ket{\Psi(0)}$ belongs to the ground state manifold before the quench dynamics. Therefore, when dynamical singularities are present, we observe a nonanalytic behavior of the rate function~\cite{Zeng2023}. \par 

In many situations, it is challenging to prepare and control a system with a large number of degrees of freedom; and therefore, experiments with small quantum systems~\cite{Jurcevic2017,Chen2020} or quantum computers implementations~\cite{Pomarico2023} are designed to test dynamical singularities. Technically, the LE usually has no exact zeros for finite-size systems, except for fine-tuned post-quench parameters, which fulfill specific constraint conditions depending on the physical implementation~\cite{Zeng2023}. Generally, when $N \rightarrow \infty$, the zeros of the LE approach to the real-time axis for quenches crossing the quantum phase transition point~\cite{Heyl2013}. A natural question arises: how do we predict or classify dynamical singularities of the rate function in terms of controllable physical parameters in a finite-size system? This question is relevant since, as we discuss later, the existence of dynamical singularities is connected with a dynamical symmetry recovery of the ground state manifold. \par

One of the most studied models in physics is the transverse field Ising model (TFIM)~\cite{Heyl2013,Jurcevic2017}, where the external magnetic field is perpendicular to the interaction axis. The TFIM can be analytically solved, allowing analytical expressions for the critical times $t_n^{\ast}$ such that $\mathcal{L}(t_n^{\ast}) = 0$ in the thermodynamical limit~\cite{Heyl2013}. Recent theoretical and numerical simulations in color centers in diamond show that including a magnetic field with transverse and longitudinal components leads to exciting applications of dynamical singularities for quantum metrology~\cite{Raul2022}, even in the thermodynamical limit. Thus, going beyond the TFIM opens the door to exploring applications that use dynamical singularities as a quantum resource. \par

Quantum machine learning (QML) is a promising research area combining quantum/classical algorithms and data to solve highly complex problems for classification and regression~\cite{qml}. To obtain a quantum advantage using QML, a feature map with quantum circuits has been proposed, mainly by exploiting unique quantum effects, such as superposition or entanglement~\cite{kernel2}. Also, a recipe of quantum algorithms has been used to study phase transition in many-body systems~\cite{qkernel2}, non-Markovianity~\cite{qkernel1}, among others. In particular, quantum Kernels are a pivotal base in some QML methods~\cite{book1}, where their effectiveness has been studied and widely compared with other techniques in QML~\cite{qml1}. For example, the superiority of quantum Kernel methods for quantum phase recognition in many-particle quantum systems has been studied recently~\cite{lo-qpr}. Among the main advantages of using quantum Kernels, we found the easy implementation in known algorithms that allow the Kernel trick, such as the Support Vector Machine (SVM) algorithm~\cite{book1}. Inspired by the original idea of using quantum Kernels to classify quantum phase transitions of matter using variational circuits~\cite{qkernel2}, we focus on a hybrid SVM algorithm, where quantum states are used to define a Kernel function, and a classical SVM is applied to the binary classification of dynamical singularities. \par

In this work, we present a finite-sized system of $N$ long-range interacting qubits, where dynamical singularities of the rate function are manipulated through external fields. We demonstrate that using the information encapsulated in the ground state and the one-tomography state of the system (wave function or density matrix), one can formulate two physically motivated quantum Kernels to classify dynamical singularities efficiently. We also demonstrate that our approach can be adapted to the effect of time-dependent Hamiltonians and Markovian baths in the weak coupling regime. Finally, we test our method by increasing the number of qubits.

\section{Quench dynamics and dynamical singularities of the rate function} \label{Model}
Let us consider the following multiqubit Hamiltonian with long-range interactions

\begin{equation}\label{Hamiltonian}
  H_{\rm s}=  H_0 + H_1 = -\sum_{i \neq j}J_{ij} \sigma_x^i \sigma_x^j - \bi{h} \cdot \sum_{i=1}^{N} \boldsymbol{\sigma}_i,
\end{equation}

where $J_{ij} = J |i-j|^{-\alpha}/N_{\alpha}$ is the long-range interaction coupling with a power-law decay $\alpha>0$, $N_{\alpha} = \sum_{i\neq j}|i-j|^{-\alpha}/N$ is a normalization factor, $\bi{h} = (h_x,h_y,h_z)$ is a generic magnetic field, and $\boldsymbol{\sigma}_i = (\sigma_x^i,\sigma_y^i,\sigma_z^i)$ is a vector containing the Pauli matrices for $S=1/2$. When $h_x = h_y = 0$ we recover the TFIM described in Refs.~\cite{Jurcevic2017,Heyl2013}. Therefore, considering a general magnetic field, we will face a more challenging scenario to analyze and classify the dynamical singularities of the rate function. \par 

\begin{figure}[ht!]
\centering
\includegraphics[width=1 \linewidth]{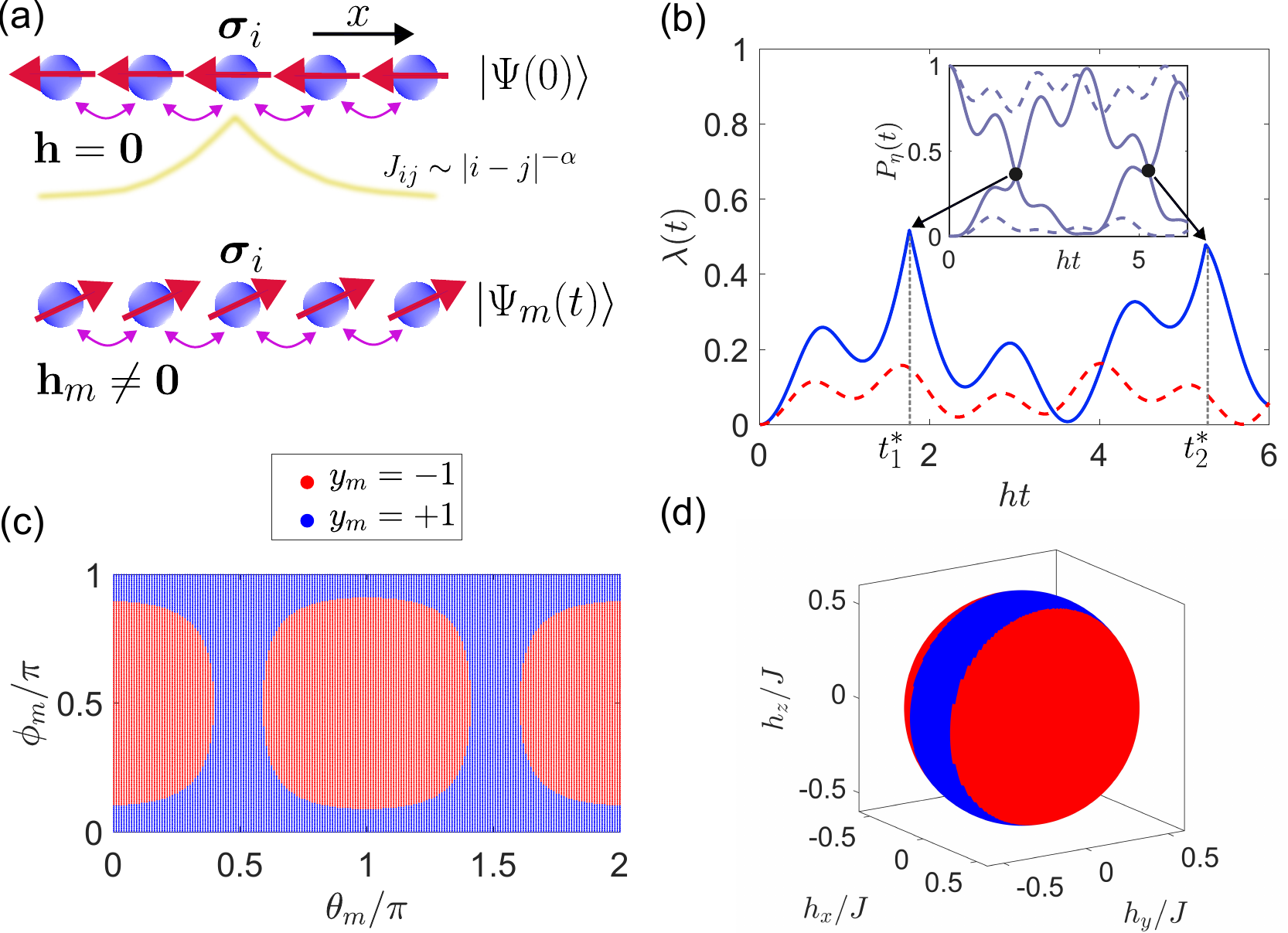}
\caption{(a) Diagram of the long-range $N$ interacting qubits illustrating the initial $\ket{\Psi(0)}$ and evolved $\ket{\Psi_m(t)}$ states when a magnetic field $\bi{h}_m$ is applied (quench dynamics). (b) Plot of the rate function $\lambda(t)$ and populations $P_{\eta}(t) = |\langle G_{\eta}| \Psi(t) \rangle|^2$ (inset) for two qubits, $\alpha=0.5$ and $h=0.6J$ for spherical angles (solid blue = $\{1.5\pi,\pi/2\}$, red dashed = $\{1.3\pi,\pi/2\}$). We note that the rate function has a singularity (solid line) when both populations satisfy the constraint $P_{+}(t_n^{\ast})=P_{-}(t_n^{\ast})$. (c) Contour plot of the features $y_{m}$ (blue = $+1$, red = $-1$) in terms of the spherical angles $\{\theta_m \times \phi_m\}$ for $200\times 200$ data points. (d) Binary sphere representation of two interacting qubits showing the magnetic field regions where dynamical singularities of the rate function are present.} 
\label{fig:Figure1}
\end{figure}

Initially, we consider $|\bi{h}| = 0$ on equation~(\ref{Hamiltonian}), and we fixed the initial many-body state as $|\Psi(0)\rangle = |G_-\rangle$ to be the one of the ground states of $H_0$, see figure~\ref{fig:Figure1}(a). We remark that $H_0$ has two degenerate ground states given by $|G_{\eta} \rangle = \prod_{j=1}^{N}|\eta\rangle_j^{\otimes}$, where $|\eta \rangle_j = (1 / \sqrt{2})(\ket{\uparrow}_j + i\eta \ket{\downarrow}_j)$ with $\eta = \pm$. The initial Hamiltonian has the property $[O,H_0] = 0$, where $O = \prod_{j=1}^{N}(\sigma_z^{j})^{\otimes}$ is the parity operator. We note that both states $|G_{\eta}\rangle$ break the $\mathds{Z}_2$ symmetry, and one interesting question is how to recover the initial symmetry of the ground state by a quantum quench process induced by a sudden change of the magnetic field?  \par 

Physically, the many-body states $\ket{\Uparrow} = \prod_{j=1}^{N}\ket{\uparrow}_j^{\otimes}$ and $\ket{\Downarrow} = \prod_{j=1}^{N}\ket{\downarrow }_j^{\otimes}$ are eigenstates of the parity operator, \textit{i.e.} $O \ket{\Uparrow}  = +\ket{\Uparrow}$ and $O \ket{\Downarrow}  = -\ket{\Downarrow}$. We remark that other quantum states, such as antiferromagnetic configurations of the form $\ket{\downarrow}_1 \otimes \ket{\uparrow}_2 \otimes \ket{\downarrow}_3 \otimes ... \otimes \ket{\uparrow}_N$ are also eigenstates of the local parity operator. However, they do not belong to the ground state manifold of $H_0$. We define $P_{\eta}(t) = |\langle G_{\eta}| \Psi(t) \rangle|^2$ as the probability of each degenerate ground state $\ket{G_{\eta}}$ with respect to the evolved state $|\Psi(t)\rangle = U(t)|\Psi(0)\rangle$, where $U(t) = \mbox{exp}(-i H_{\rm s}(\bi{h}) t)$ is the unitary evolution operator for $|\bi{h}| \neq 0$ (quench dynamics). Thus, the condition $P_{+}(t) = P_-(t)$ implies that the ground state manifold will be described by a state of the form $|G\rangle = (1/\sqrt{2})(|G_+\rangle \pm |G_-\rangle ) = \{\ket{\Uparrow} \; \mbox{or} \; \ket{\Downarrow} \}$. The latter is a condition for dynamical symmetry recovery induced by an external magnetic field. \par 

For a finite-sized quantum system with a degenerate ground state, it is usual to analyze the behavior of the rate function~\cite{Jurcevic2017} 

\begin{equation}
    \lambda(t) = \min_{\eta \in \pm}\left \{ -(1/N)\log[P_{\eta}(t)]\right\}. \label{RateFunction}
\end{equation}

Our system will exhibit dynamical singularities of the rate function when $\lambda(t)$ presents a nonanalytic behavior at some critical times $t_n^{\ast}$ such that $\lambda(t) - \lambda(t_n^{\ast}) \approx |t-t_n^{\ast}|$ in the vicinity of $t_n^{\ast}$, see solid line curve in figure~\ref{fig:Figure1}(b). Theoretically, such derivative discontinuities are given by $\lim_{\varepsilon \rightarrow 0}[\dot{\lambda}(t_n^{\ast}+\varepsilon)-\dot{\lambda}(t_n^{\ast}-\varepsilon)] = -(1/N)\Delta P_n/P_n$ with $\Delta P_n = \dot{P}_+(t_{n}^{\ast})-\dot{P}_-(t_{n}^{\ast})$ and $P_n = P_{\pm}(t_n^{\ast})$. Note that $P_-(0)=1$ and $P_+(0) = 0$, therefore the first crossing between probabilities $P_{\eta}(t)$ requires that $\dot{P}_+(t_{n}^{\ast})>0$ ($P_+$ increases) and $\dot{P}_-(t_{n}^{\ast})<0$ ($P_-$ decreases) leading to $\Delta P_n \neq 0$ (first derivative discontinuity). When $P_{+}(t_n^{\ast}) = P_{-}(t_n^{\ast})$ the ground state recovers the $\mathds{Z}_2$ symmetry and also gives us a numerically accessible criterion to determine if the system undergoes a dynamical singularity. As demonstrated in Ref.~\cite{Jurcevic2017}, in the transverse field Ising model, the critical time can be estimated as $\tau_{\rm crit} \approx \pi/4 + D_{\rm crit}(J/ h)^2$. Therefore, in our numerical simulations, we will explore the dynamics in the time window $t \in [0, T]$ such that $T \gtrsim 10^2\tau_{\rm crit}$ to examine if dynamical singularities are present.

\par

In what follows, we will explain how to generate data for further implementations of the SVM algorithm, revealing the complexity of this classification problem in terms of the angular distribution of the external magnetic field.

\section{Initial data for support vector machine}\label{DataSet}

As a proof of principles of our quantum SVM algorithm, we define our features as the set of vectors $\bi{x}_m = (\theta_m,\phi_m)$ for a given value of $\alpha > 0$ and $h$, where $m = 1,2,..,M$ indicates the dimension of the initial dataset. We use spherical coordinates for the magnetic field $\bi{h} = (h\sin\phi\cos\theta, h\sin\phi\sin\theta, h\cos\phi)$, where $h>0$, $\theta \in [0, 2 \pi]$, and $\phi \in [0,\pi]$. Physically, we are classifying dynamical singularities of the rate function in terms of the angular orientation of the magnetic field for a system with $N$ interacting qubits. Later, in Section~\ref{Results2}, we will add the effect of the magnitude of the magnetic field. As mentioned, we focused on systems with some defined power-law decay $\alpha$ for the long-range interaction. 
We remark that $0<\alpha<3$ in diverse realizations of dynamical quantum phase transitions using trapped ions~\cite{Jurcevic2017}. Thus, we fix $\alpha = 0.5$, but the following analysis and results are not affected by this choice. As labels, we use the binary variable $y_m = \{+1,-1\}$, where $+1$ ($-1$) means that our system does (does not) have dynamical singularities in the rate function. We numerically generate our features $(\theta_m,\phi_m)$ such that $\theta_m \in [0, 2\pi]$ and $\phi_m \in [0, \pi]$ are equally distributed points (rectangular grid) in their respective domains. Using numerical simulations based on the Hamiltonian~(\ref{Hamiltonian}) and applying the criteria $P_{+}(t_n^{\ast}) = P_{-}(t_n^{\ast})$, we determine the labels in our model (the binary value of $y_m$). Here, $t_n^{\ast}$ is a critical time when the system exhibits a non-analytical behavior in the rate function. In figure~\ref{fig:Figure1}(c), we plot the feature $y_m$ in terms of the spherical angles $(\theta_m, \phi_m)$ for the particular case of $N=2$ interacting qubits with $h =  0.6 J$ and $\alpha=0.5$. \par 

\begin{figure}[ht!]
\centering
\includegraphics[width=1 \linewidth]{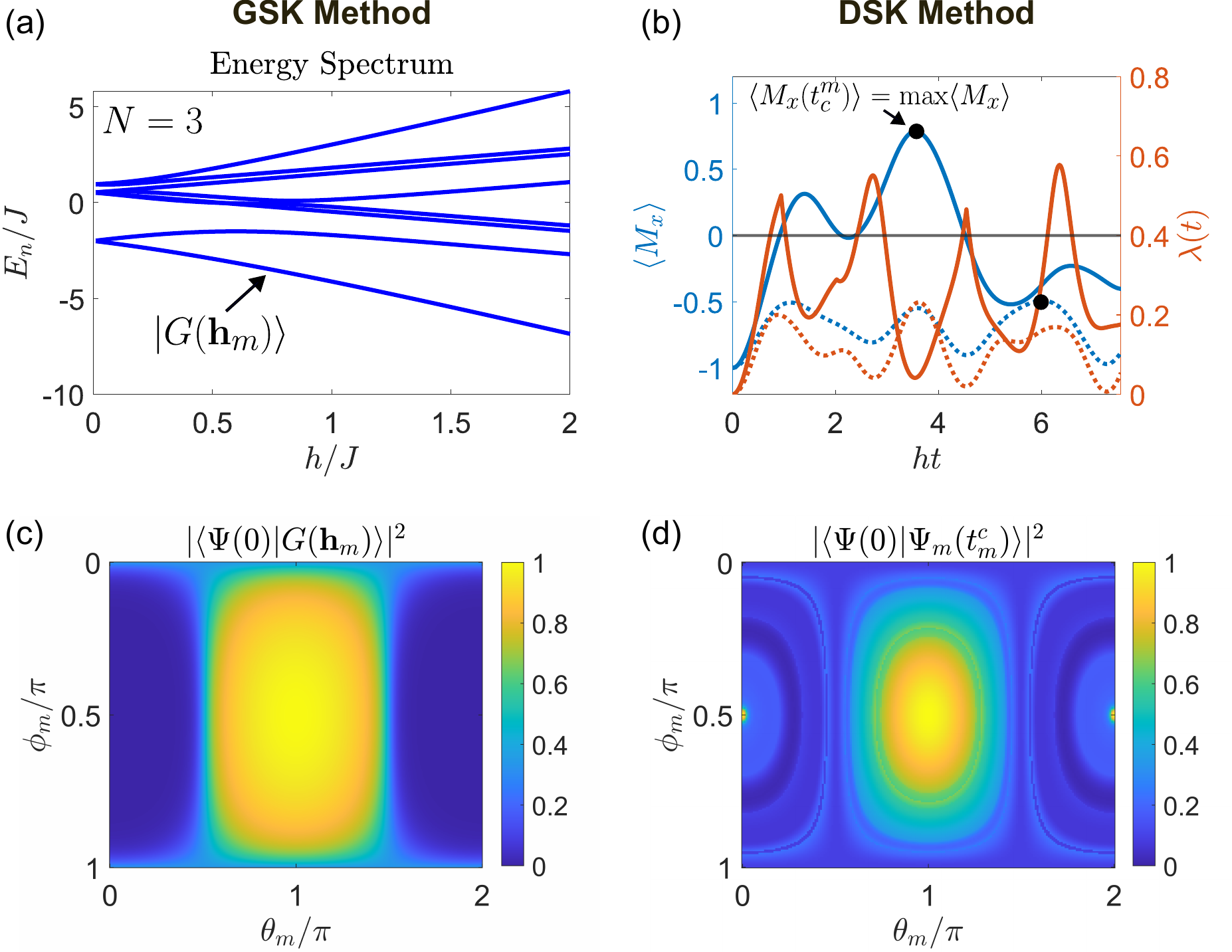}
\caption{(a) Energy spectrum of the system for $N=3$ qubits, illustrating the ground state $\ket{G(\bi{h}_m)}$ used for the GSK method. (b) Dynamical behavior of the magnetization component $\langle M_x \rangle$ and rate function $\lambda(t)$ for two different cases (solid line is for $y_m = +1$ and dashed line of for $y_m = -1$). The critical time $t_c^m$ gives the maximum $\langle M_x \rangle$ value. Probabilities (c) $|\langle \Psi(0)|G(\bi{h}_m) \rangle|^2$ and (d) $|\langle \Psi(0)|\Psi_m(t_m^c) \rangle|^2$ in terms of the magnetization angles $\theta_m$ and $\phi_m$.} 
\label{fig:Figure2}
\end{figure}

Moreover, we introduce a binary sphere representation to illustrate the magnetic field regions where dynamical singularities are (are not) present. In figure~\ref{fig:Figure1}(d), we define a colored sphere surface with a radius equal to $h$ such that regions in blue (red) are associated with the label $y_m = +1$ ($y_m=-1$). We remark that Figs.~\ref{fig:Figure1}(c)-(d) correspond to the same dataset. However, we only show the symmetry presented in our dataset in figure~\ref{fig:Figure1}(d). We observe that for a given magnetic field with a label $y_m$, any magnetic field rotated around the $y$-axis preserves the same label. In other words, if we apply a quench dynamics using the magnetic field $\bi{h}$, then the rotated field $\bi{h}' = R_y(\beta) \bi{h}$ has the same label $y_m$, with $\beta \in [0, 2\pi]$ representing an angle and $R_y(\beta)$ describing the rotation matrix around the $y$-axis. However, as we shall discuss later, this symmetry is broken when time-dependent fields or Markovian losses are present in our system. In what follows, we will motivate using Kernels in machine learning by transitioning from classical to quantum Kernels. 

\section{From Classical to Quantum Kernels}

Given the basic notion of data in this problem, we now shall introduce the concept of classical Kernel. Let us suppose that we have two different vectors $\bi{x}_m, \bi{x}_{m'} \in \mathds{R}^{M}$ representing features of a dataset, and we are interested in defining a similarity measure for such vectors, where $M$ is the dimension of our dataset. We define the Kernel function $K(\bi{x}_m,\bi{x}_{m'}): \mathds{R}^M \times \mathds{R}^M  \mapsto \mathds{R}$ such that $0 \leq K(\bi{x}_m,\bi{x}_{m'}) \leq 1$ represent a normalized measure between features. In this case, we deal with two extreme cases, namely $K(\bi{x}_m,\bi{x}_{m'})=1$ (same data, $\bi{x}_m = \bi{x}_{m'}$) and $K(\bi{x}_m,\bi{x}_{m'})=0$ (totally different data). For simplicity, we use the notation $K_{mm'} = K(\bi{x}_m,\bi{x}_{m'})$ to denote the Gram matrix elements associated with the Kernel function. We remark that $K_{mm'}$ is a positive semi-definite matrix satisfying: i) $K_{mm} = 1$, ii) $K_{mm'} = K_{m'm}$, and iii) $0 \leq K_{mm'} \leq 1$. Support Vector Machine algorithms require this Kernel function to efficiently map initial input data into a high-dimensional feature space (feature map), which allows us to improve nonlinear classification algorithms (\an{see~\ref{AppendixA}} for further technical details). \par

The basic idea behind quantum Kernels is to map initial input data into quantum states and then explore quantum metrics to introduce a suitable and well-behaved quantum Kernel function. The latter leads to the Quantum Support Vector Machine (QSVM) paradigm, where quantum states carry information on the features and are used to define quantum kernels. In this work, we use a classical dataset $\{\bi{x}_m\}$ to build our quantum dataset $\{\ket{\Psi_m}\}$ (wave function) or $\rho_m$ (density matrix) by exploiting the rules of quantum mechanics (evolution and ground state) of this particular system. Nowadays, the feature map can be combined with parameterized gates in variational quantum circuits~\cite{vqc1, vqc2} or data re-uploading algorithms~\cite{datareup}. Quantum Kernel methods can take some advantage of metrics of quantum states obtained from the feature map for each data~\cite{kernel1}. Throughout this work, two kinds of Kernels are explored to analyze the performance of our algorithm:

\begin{eqnarray}
    K^{\rm qlin}_{mm'} &=& \bi{I}^{\rm Q}_{mm'}, \label{Qlin}\\
    K^{\rm qrbf}_{mm'} &=& \exp\left(-\gamma\sqrt{1-\bi{I}^{\rm Q}_{mm'} }\right), \label{Qrbf} 
\end{eqnarray}

where $\bi{I}^{\rm Q}_{mm'}$ is the quantum inner product between two quantum states, $\gamma>0$ is a free parameter and ``qlin'' and ``qrbf'' stands for quantum versions of the linear and radial basis function, respectively. Here, we must clarify that the quantum inner product depends on the nature of the problem (closed or open quantum system). For a closed quantum system modeled by the Schr\"{o}dinger equation and pure states, we use the overlap over between quantum states $\bi{I}^{\rm Q}_{mm'}=|\langle{\Psi_m}|\Psi_{m'}\rangle|^2$~\cite{Buhrman2021}, where $\ket{\Psi_{m}}$ is a wave function that encapsulated information of the feature $\bi{x}_m$. We will discuss two different methods in Sections~\ref{SectionGSK} and~\ref{SectionDSK}. Instead, for an open quantum system, we use the overlap for mixed states, which is defined as $\bi{I}^{\rm Q}_{mm'}= \mbox{Tr}(\rho_m \rho_{m'})$~\cite{overlap}, where $\rho_m$ is a density matrix that properly carries information about the feature $\bi{x}_m$.\par

To obtain a quantum advantage with quantum Kernels over classical approaches, one can explore unique quantum effects such as superposition or entanglement~\cite{kernel2}. It has been demonstrated that one can engineer datasets to outperform classical methods with quantum Kernel methods~\cite{dataqml}, showing the strong influence of data and the chosen feature map, in line with Ref.\cite{kernel3}, where quantum advantage in a real-world problem has been found. Also, in the quantum computing paradigm, it is possible to obtain the inner product of quantum data using known circuits based on the overlap between quantum states, like the swap test or Bell basis algorithm~\cite{overlap} or the inversion test~\cite{inversion}. Therefore, it is possible to calculate the quantum Kernel with quantum data, as in Refs.~\cite{qkernel1, qkernel2}. Here, we use classical data (magnetic field) to create quantum states (quantum data). Then, we propose a QSVM method to classify the dynamical singularities of the rate function. Now, we will focus on how to calculate the inner product $\bi{I}^{\rm Q}_{mm'}$ for the closed case inspired by physical arguments.

\subsection{Ground State Kernel (GSK)} \label{SectionGSK}

Dynamical singularities of the rate functions are connected to critical phenomena related to the behavior of the ground state manifold when a quench dynamics is present. Thus, proposing a quantum Kernel based on the information stored in the ground state for a given set of physical parameters emerges naturally. It is worth noticing that the connection between the overlap of ground states with different external parameters and quantum phase transition (QPT) has been studied in Ref.~\cite{qptoverlap}, observing that the inner product tends to decrease significantly in the thermodynamical limit. This behavior has been related to the phenomenon known as the Anderson orthogonality catastrophe~\cite{anderson}. In addition, Ref.~\cite{qkernel2} also uses this overlap property to classify QPT using quantum Kernels. Most importantly, in Ref.~\cite{qptoverlap}, the authors find a connection between the overlap and decay in the Loschmidt echo, showing that the inner product of ground states can capture information about dynamical properties. Motivated by these physical observations and rigorous methods, we first introduce the quantum data as $\ket{\Psi_m} = \ket{G(\bi{h}_m)}$, where $\ket{G(\bi{h}_m)}$ is the ground state of $H_{\rm s}(\bi{h}_m)$ (for $|\bi{h}_m| \neq 0$), as shown in figure~\ref{fig:Figure2}(a). Here, we are using the notation $\bi{h}_m$ to give a more physically intuitive notion of the role of an arbitrary magnetic field $\bi{h}_m$ viewed as a feature (or $\bi{x}_m$ as explained in Section~\ref{DataSet}). We call this method Ground State Kernel (GSK), where the inner product is defined as 

\begin{equation} \label{GSK}
\bi{I}^{\rm GSK}_{mm'} = |\langle G(\bi{h}_m)|G(\bi{h}_{m'}) \rangle|^2,
\end{equation}

where $H_{\rm s} \ket{G(\bi{h}_m)} = E_g(\bi{h}_m)\ket{G(\bi{h}_m)}$, with $E_g(\bi{h}_m)$ being the ground state of $H_{\rm s}$ for the magnetic field $\bi{h}_m$. Note that this ground state is calculated considering the magnetic field used in the quench dynamics. Thus, we use equation~(\ref{GSK}) for the calculation of $K^{\rm qlin}_{mm'}$ and $K^{\rm qrbf}_{mm'}$, respectively. Naturally, the ground state $\ket{G(\bi{h}_m)}$ is given by a linear combination of the multiqubit basis $\ket{n} = \prod_{j=1}^{N}\ket{\varphi}_j^{\otimes}$ with $\ket{\varphi}_j \in \{\ket{\uparrow}, \ket{\downarrow}\}$. Thus, $\ket{G(\bi{h}_m)} = \sum_{n=1}^{2^{N}} c_n\ket{n}$, where the probability amplitude coefficients $c_n = c_n(\bi{h}_m)$ encodes information about the external magnetic field. However, because of the normalization constraint of the wave function, each amplitude coefficient $c_n$ will carry less information when the system size grows ($N \rightarrow \infty$). This will be crucial to understanding the limitation (or less accuracy) of this method for many qubits. In figure~\ref{fig:Figure2}(c), we plot the probability $|\langle \Psi(0)| G(\bi{h}_m)\rangle|^2$ to illustrate that the symmetrical behavior resembles a good analogy to the feature-label map given in figure~\ref{fig:Figure1}(c). This simple and intuitive inspection tells us that using the ground state (with $|\bi{h}_m| \neq  0$) could be enough, in principle, to capture the essence of the classification problem, similar to the amplitude embedding-like parametrization approach~\cite{Grant2019}. It is important to note that this quantum Kernel can be implemented in a quantum computer. In this case, we need to obtain a quantum circuit for obtaining the groundstate of $H_{s}(\bi{h}_m)$, a task realizable with variational quantum eigensolver, like the algorithms proposed in Ref.~\cite{qkernel2}.

\subsection{Dynamical State Kernel (DSK)} \label{SectionDSK}

Regarding the relevance of the ground state manifold in the foundations of critical phenomena, it is intuitive that some additional information can be taken from the evolution. A dynamical singularity is a process where the system tries to recover the initially broken symmetry. Thus, adding some information about the instantaneous wave function (or density matrix) could improve, in principle, our strategy for classification. The problem here is how to have a full tomography of the quantum state. Our approach here will focus on a single state tomography of the wave function (or density matrix). Fundamentally, one can introduce a proper order parameter to characterize the dynamical singularities of the rate function. In Ref~\cite{Jurcevic2017}, it is observed that a change of sign in the macroscopic magnetization $\langle M_x \rangle$ occurs in particular times when dynamical singularities of the rate function are present. Then, inspired by these observations, we introduce the Dynamical State Kernel (DSK) method by using the following inner product 

\begin{equation}\label{DSK}
\bi{I}^{\rm DSK}_{mm'} = |\langle \Psi_m(t_{m}^c)|\Psi_{m'}(t_{m'}^c)\rangle|^2.    
\end{equation}

Here, we include $t_m^{\rm c}$ as an additional free parameter (to be found), which only considers a single state tomography of the wave function. Based on this, our quantum data is given by $\ket{\Psi_m(t_m^{\rm c})}=\exp(-iH_{\rm s}(\bi{h}_m)t_m^{\rm c})\ket{\Psi(0)}$. We numerically determine the critical time $t_m^{\rm c}$ as the time required to reach the maximum value of $\langle M_x \rangle = (1/N)\sum_{i=1}^{N}\langle \sigma_x^i\rangle$ in a large time window where dynamical singularities can be observed (if it is present). Note that the choice of $\langle M_x \rangle$ over $\langle M_y \rangle$ or $\langle M_z \rangle$ is because of symmetry arguments and depends crucially on what kind of interaction we have in our original Hamiltonian~(\ref{Hamiltonian}). In figure~\ref{fig:Figure2}(b), we show the magnetization and rate function for three interacting qubits with $h = 1.2J$, $\alpha = 0.5$, $(\theta,\phi) = (1.5\pi, \pi/2)$ (solid line), and $(\theta,\phi) = (1.3\pi, \pi/2)$ (dashed line). We note that $\langle M_x(t_c^{m})\rangle >0$ when a critical behavior of the rate function is present, otherwise $\langle M_x(t_c^{m})\rangle <0$. \par

It is worth noticing that this methodology can be implemented in a quantum computer using digital quantum simulation. The latter can be done using the usual trotterization process of the time evolution operator in the form $\ket{\Psi_m(t_m^c)}= U(\bi{h}_m,t_m^c)\ket{0}^{\otimes N}$, where $U(\bi{h}_m,t_m^c)$ is a set of quantum gates. This results in the trotterization of the time evolution operator $\exp(-iH_{\rm s}(\bi{h}_m)t_m)$ (\an{see~\ref{AppendixB}} for specific details). We remark that a complete quantum algorithm is complicated when the number of qubits increases since one requires multiple quantum gates to perform the swap test necessary for the quantum state tomography. Now, we have the main ingredients to show the power of the QSVM algorithm using both GSK and DSK methods. 

\begin{figure}[ht!]
\centering
\includegraphics[width=1 \linewidth]{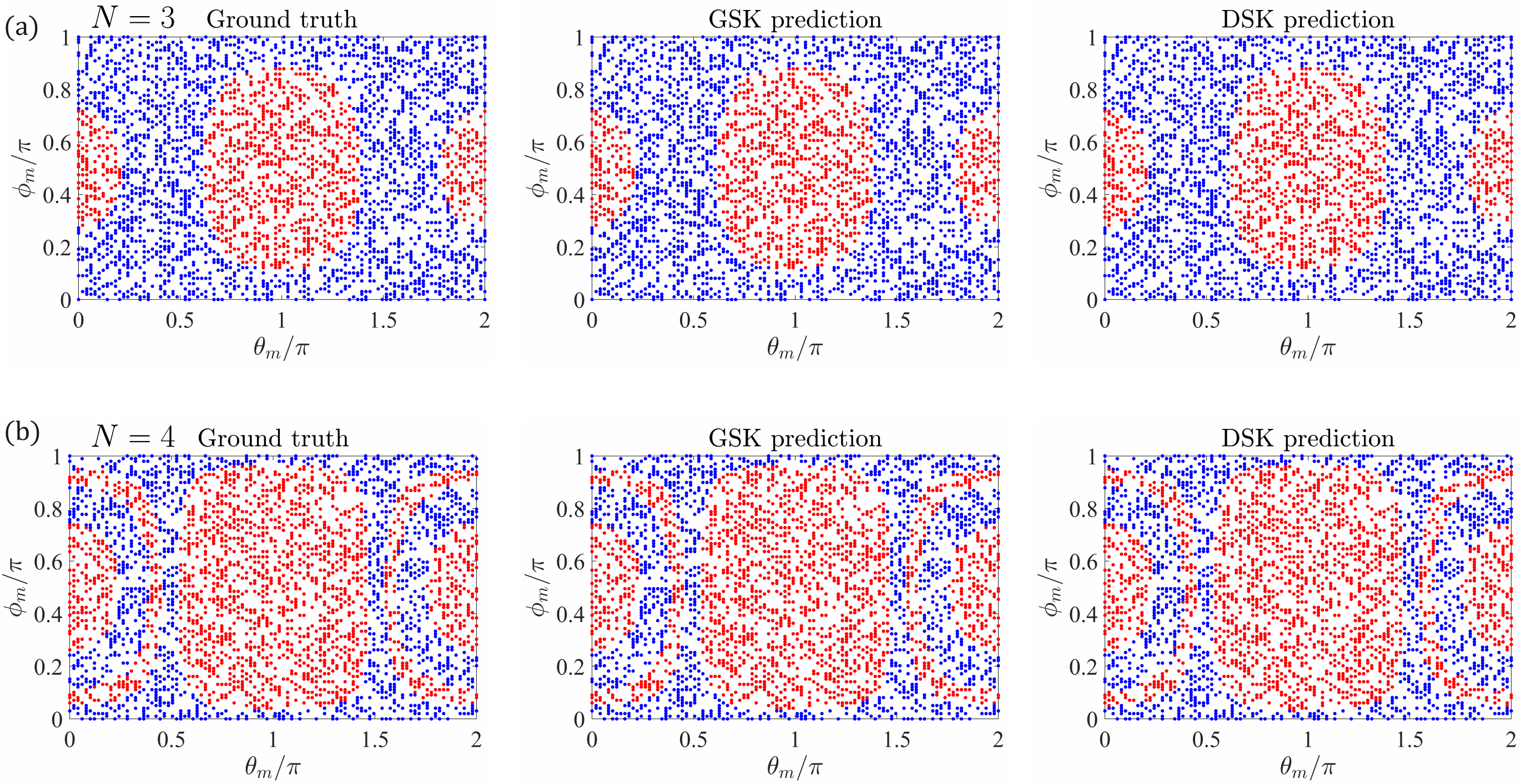}
\caption{(a) Ground truth and GSK and DSK predictions for $N=3$ qubits using $h=1.2J$. (b) Ground truth, GSK, and DSK predictions for $N=4$ qubits using $h=0.8J$. In all simulation we use $\alpha = 0.5$ and the initial condition $\ket{\Psi(0)} = |G_-\rangle$.} 
\label{fig:Figure3}
\end{figure}

\section{Results}

In this Section, we discuss the main results of this work for the accuracy and adaptability of our QSVM algorithm. In particular, we first focus on the accuracy of both GSK and DSK methods compared to a classical SVM classifier. Then, we introduce a more challenging scenario by including the effect of a time-dependent transverse field and losses modeled by a Markovian master equation in the Lindblad form. Finally, we analyze the impact of increasing the number of qubits on the proposed method.

\subsection{Accuracy of GSK and DSK methods} \label{Results1}

To analyze the accuracy of the quantum Kernels, we focus on two cases where the data distribution exhibits different geometrical patterns. We construct our initial classical dataset using an equally distributed grid of $100 \times 100 = 10^4$ points for the magnetization angles. The $100 \times 100$ 2D discretization is chosen to fulfill two relevant criteria: (i) angular precision in regions with different labels and (ii) enough amount of data points to train our SVM algorithm. Thus, $\bi{x}_m = (\theta_m, \phi_m)$ is the vector describing our features (with a fixed magnetic field amplitude $h$). The labels $y_m$ are computed numerically, where we use the physical criteria explained in Sec.~\ref{Model}, as shown in figure~\ref{fig:Figure1}(c). We chose 70\% of this initial dataset for training and the remaining 30\% for testing. We remark that the training and testing datasets are disjoint. Therefore, we test our predictions using data we never used in the training step. The training data is randomly chosen during all simulations. The best predictions are obtained with $K_{mm'}^{\rm qrbf}$ for GSK (with $\gamma=1$) and $K_{mm'}^{\rm qlin}$ for DSK. \par 

In figure~\ref{fig:Figure3}, we show the predictions of the GSK and DSK methods compared with the ground truth (testing data for $N=3$ and $N=4$). For $N=3$ and $h=1.2J$, we found a symmetrical distribution of red ($y_m = +1$) and blue ($y_m=-1$) regions with small spherical regions at angles $\theta_m \approx 0$ and $\theta_m \approx 2\pi$, see figure~\ref{fig:Figure3}(a) (left figure). In this case, we reach accuracies equal to 99.9\% (GSK) and 99.7\% (DSK). We remark that, as shown in figure~\ref{fig:Figure2}(c) and (d), both quantum states $\ket{G(\bi{h}_m)}$ (GSK) and $\ket{\Psi_m(t_m^c)}$ (DSK) encodes valuable information about the angular distribution for this classification problem. However, even in a more challenging scenario in terms of angular distribution, as shown in figure~\ref{fig:Figure3}(b), where $N=4$ and $h=0.8J$, we obtain 98.7\% (GSK) and 98.4\% (DSK) accuracies. \par

To have a fair comparison, we solve the same classification problem using the classical SVM algorithm, implemented in MATLAB by applying the automatic hyperparameter optimization~\cite{SM1}. Using the Support Vector Classifier method, we find accuracies around $99.8\%$ ($N=3$) and $98.2\%$ ($N=4$) after applying the hyperparameter optimization using cross-validation. These results illustrate that quantum kernel methods allow a precision comparable to the classical version with a small statistical improvement but not significant. At this stage, the most remarkable observation is that our GSK method outperforms the DSK and classical SVC methods. Surprisingly, GSK does not require any evolution to solve this classification problem, even though the problem itself is dynamical. In other words, the information about $(\theta_m,\phi_m)$ is faithfully encoded by the ground state $\ket{G(\bi{h}_m)}$, giving us the benefit of solving the classification problem without the necessity of solving (or simulating) the dynamics of the multiqubit system. However, suppose the problem has some dynamical features not explicitly given in the ground state. In that case, the DSK method will be, in principle, the best option. The latter can occur, for instance, when we deal with dissipation, stochastic noise, or time-dependent Hamiltonians. In the following sections, we will study the effect of dissipation and time-dependent fields to show the validity and robustness of our QSVM scheme. \par 

\begin{figure}[ht!]
\centering
\includegraphics[width=1\linewidth]{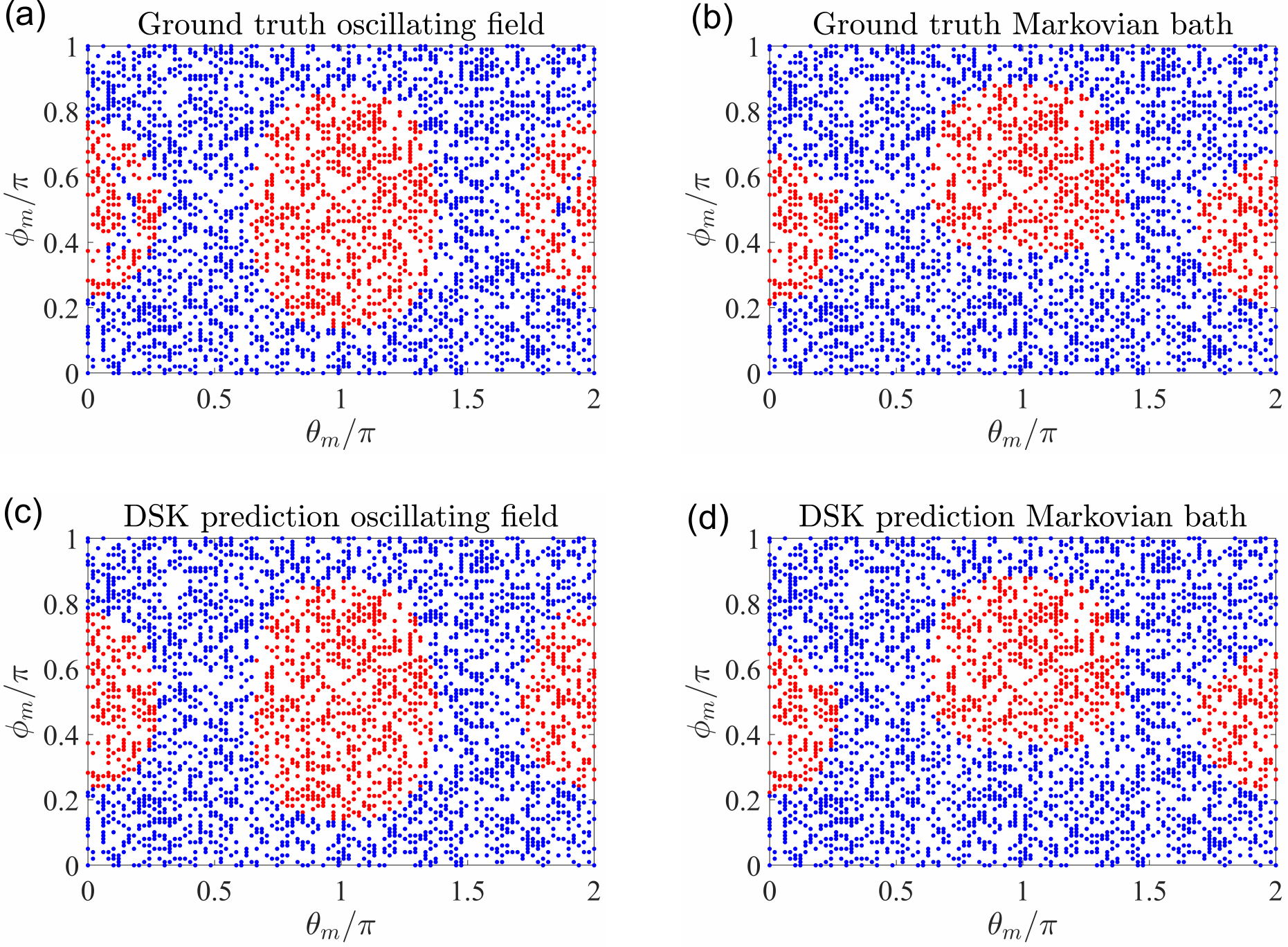}
\caption{Ground truth (testing dataset) for the oscillating magnetic field (a) and Markovian master equation (b). Predictions for the oscillating magnetic field (c) and Markovian master equation (d) using the DSK method. For the oscillating field $B_z(t) = B_{z0} \sin(\omega t)$, we use $B_{z0} = 0.05h$, $w = 1.2J$ for $N=2$ qubits, where \an{$h = 0.95J$} and $\alpha = 0.5$. For the open case, we use damping rates $\gamma = 0.02 J$, using the same system parameters that cases (a) and (c).} 
\label{fig:Figure4}
\end{figure}

\subsection{Effect of time-dependent Hamiltonians} \label{Results3}

One can steer the system's dynamics in many physical situations to introduce and explore some natural responses of the system. In particular, time-dependent sinusoidal fields are very useful in electron spin resonance experiments~\cite{NJPAriel2020}, Floquet engineering~\cite{PRBGuillermo2022}, and also in recent theoretical studies of dynamical quantum phase transition~\cite{Raul2022}. In this direction, we can add a time-dependent driving field to solve a more subtle question. Can our QSVM algorithm classify dynamical singularities under dynamical control? To explore this question, we consider the time-dependent Hamiltonian
\begin{equation}
    H(t) = H_{\rm s} + B_z(t) \sum_{i=1}^{N} \sigma_z^{i},
\end{equation}
where $H_{\rm s}$ is the Hamiltonian introduced in equation~(\ref{Hamiltonian}), and $B_z(t) = B_{z0} \sin(\omega t)$ is a magnetic field that naturally induces rotations on the initial state $\ket{\Psi(0)} = |G_-\rangle = \prod_{j=1}^{N}|-\rangle_j^{\otimes}$. Here, we introduce the amplitude $B_{z0}$ and frequency $\omega$ as new parameters to test the system's response. Then, since a time-dependent signal is helping to rotate the initial ground state, the prediction of singularities in $\lambda(t)$ becomes more challenging than a purely time-independent Hamiltonian evolution (as we discussed in Section~\ref{Results1}). Moreover, in the regime $B_{z0} \ll h$, we are physically analyzing the response of the system when magnetic field fluctuations are incorporated into the $z$-axis. We remark that at $t=0$, $H(t=0) = H_{\rm s}$, and therefore, no information about $B_{z0}$ and $\omega$ can be obtained from the initial ground state. \par

As an example, in figure~\ref{fig:Figure4}(a), we observe the pattern (ground truth) for the classification of dynamical singularities using $B_{z0} = 0.05h$ and $w = 1.2J$ for $N=2$ qubits. We observe that by including an oscillating field, the patterns change in comparison with $N=2$ (see figure~\ref{fig:Figure1}(c)) (even for a small value of $B_{z0}$), and a small blue ring appears around $\theta_m \approx 0.2 \pi$ and $\theta_m \approx 1.8 \pi$. For this classification problem, we obtain similar accuracies for the DSK ($97.8\%$) and  GSK ($97.4 \%$) methods, which reveals that our approach can also deal with time-dependent Hamiltonians when the external driving has a relatively small strength (weak-coupling). The predicted data distribution is shown figure~\ref{fig:Figure4}(c), where, for simplicity, only DSK predictions are plotted (since they look similar to GSK predictions). During our simulations, we noted that new features (like the tiny blue rings) arising from the $B_z(t)$ effect are the most difficult to predict when the amplitude $B_{z0}$ is comparable with $h$, but DSK always gives better results. The latter allows us to answer one fundamental question. When a time-dependent effect is not encapsulated in the ground state information (\textit{i.e.} the probability amplitudes), DSK is a better option than GSK for constructing a quantum kernel. However, GSK is the most practical since it does not require solving the dynamics, leading to an incredibly faster method and reaching good accuracies. \par

\begin{table}[ht!]
\centering
\caption{\label{Table1} The Table compares the GSK (Ground State Kernel) and DSK (Dynamical State Kernel) methods in terms of the efficiency for time-dependent Hamiltonians, Markovian noise, number of qubits, and execution time. We used different numbers of qubits and we reported the best accuracy (Accu.) using our QSVM methods for closed quench dynamics.}
 \label{table}
\begin{tabular}{ ||p{1.3cm}||p{2 cm}|p{2 cm}|p{2 cm}|p{4 cm}| p{2 cm}|}
 \hline
 \multicolumn{6}{|c|}{Comparison between GSK and DSK} \\
 \hline 
 \centering Method & \centering Time-dependent Hamiltonian  & \centering Markovian bath & \centering Advantages & \centering N$^{\circ}$ qubits &  Run time\\
 \hline
 GSK  &  \small{Good results for classification only when the initial dataset is similar to the time-independent case}  &  \small{Good results when losses are small or when dynamical singularities occur in a short time scale compared to losses} & \small{Fast and do not require time evolution to make predictions. Good results when losses and time-dependent effects are negligible} & \small{N=2, Accu. = 97.03\%  N=3, Accu. = 92.76\% 
 N=4, Accu. = 92.08\%  N=5, Accu. = 91.84\% N=6, Accu. = 92.78\%  N=7, Accu. = 92.46\% N=8, Accu. = 91.87\%} & \small{Fast method that only requires the ground state wave function for different physical parameters (minutes to hours)}\\
 \hline
 DSK &  \small{Best option and good results for classification. The method can be adapted to learn new patterns in the initial dataset} & \small{Best option when losses are small or when dynamical singularities occur in a short time scale} & \small{Excellent accuracies and can be adapted to scenarios where GKS fails} & \small{N=2, Accu. = 98.80\%  N=3, Accu. = 99.11\% 
 N=4, Accu. = 98.92\%  N=5, Accu. = 99.02\% N=6, Accu. = 99.04\%  N=7, Accu. = 98.43\% N=8, Accu. = 98.18\%} & \small{Time-consuming method that requires the calculation of the Kernel function for different physical parameters (few hours)}\\
 \hline
\end{tabular}
\end{table}

\subsection{Effect of Markovian noise} \label{Results4}

In general, a quantum system is affected by unavoidable interactions with the surrounding environment. One typically uses the Langevin approach or the theory of open quantum systems to model system-environment interactions from semi classical or purely quantum processes. Here, we adopt the usual strategy of considering a memoryless environment described by a Markovian master equation in the Lindblad form $(\hbar = 1)$

\begin{equation}\label{ME}
    \dot{\rho} = -i[H_{\rm s}, \rho] + \sum_{k=1}^{N_{\rm c}}\gamma_k \left[L_k \rho L_k^{\dagger} - {1 \over 2}\{L_k ^{\dagger}L_k, \rho\} \right]
\end{equation}

where $\rho(t)$ is the density matrix, $\gamma_k$ is the rate of the $k$th qubit, $L_k$ is the quantum operator that describes the dissipative process, and $N_{\rm c}$ is the number of dissipative channels. As a proof of principles, we consider $\gamma_k = \gamma$ and \an{$L_k = \sigma_x^k -i \sigma_y^k$}. Physically, we are modeling a common reservoir where each qubit experiences spontaneous emission with a constant (positive) rate $\gamma >0$. Beyond the Markovian and weak-coupling approximations, one can consider more sophisticated models by including strong interactions~\cite{AgarwalPRB2013,ArielPRA2020}, non-Markovian effects~\cite{ArielPRA2020}, and effective dissipative operators with exchange and on-site Lindbladians~\cite{NJPAriel2020_LL}. As the initial condition, we use $\rho(0) =  \ket{G_-}\bra{G_-}$, and the instantaneous density matrix $\rho(t)$ can be used to determine the rate function given in equation~(\ref{RateFunction}) by redefining the probabilities as $P_{\eta}(t) = \bra{G_{\eta}} \rho(t) \ket{G_{\eta}}$. \par

Let us analyze the effectiveness of both GSK and DSK methods in the open case. In this case, we use $\bi{I}^{\rm Q}_{mm'}= \mbox{Tr}(\rho_m \rho_{m'})$, where $\rho_m$ is constructed by differently depending on the method. For the GSK method, we define the inner product as $\bi{I}^{\rm GSK}_{mm'}  = \mbox{Tr}(\rho_m(G) \rho_{m'}(G))$, where $\rho_m(G) = |G(\bi{h}_m)\rangle \langle G(\bi{h}_m) |$. Similarly, for the DSK method, we use  $\bi{I}^{\rm GSK}_{mm'}  = \mbox{Tr}(\rho_m(t^c_m) \rho_{m'}(t^c_{m'}))$, where $\rho_m(t^c_m)$ is the density matrix at time $t^c_m$ such that the magnetization $\langle M_x \rangle $ is maximized (\an{see~\ref{AppendixA}}). \par 

When a spontaneous emission rate $\gamma = 0.02J$ is included per site for $N=2$ qubits, the pattern (ground truth) is slightly modified compared to the close case. In figure~\ref{fig:Figure4}(b), we note that the symmetry of the dataset around $\phi_m = \pi/2$ ($x$-axis) is broken due to this amplitude damping noise (see figure~\ref{fig:Figure1}(c) for the close case). We obtain the following accuracies for the DSK ($99.2\%$) and  GSK ($98.1 \%$) methods (see figure~\ref{fig:Figure4}(d) for DSK predictions). As discussed in the previous section, DSK is the most promising method of capturing dynamical signatures for open quantum systems. Since predictions of both methods (GSK and DSK) are reasonably in good agreement with the ground truth, we will focus now on the effectiveness of each method when the number of qubits increases. Another important dissipative channel is phase damping, which is discussed \an{in~\ref{AppendixC}}. \par

\subsection{Increasing number of qubits} \label{Results2} 

Our previous explorations focused on the classification problem for relatively small quantum systems ($N=2,3,4$) using both GSK and DSK and fixing the amplitude of the magnetic field $\bi{h}$. One crucial point is how our approach can deal with a high-dimensional Hilbert space~\cite{Albarran2022}. By increasing the number of qubits, we are dealing with computational aspects in terms of resources and physical phenomena like the Anderson orthogonality catastrophe~\cite{anderson} for the GSK method. As shown in Table~\ref{table} (see column $N^{\circ}$ of qubits), we analyze the performance of both GSK and DSK methods for $N = 2, 3, ..., 8$ qubits to illustrate the accuracy of both Kernels. The latter is relevant for quantum machine learning algorithms for the noisy intermediate-scale paradigm and for systematically understanding the thermodynamic limit in dynamical quantum phase transition. \par

In addition, to make this discussion more interesting, we add a third feature to the problem: the magnitude $h$ of the magnetic field. Now, as classical dataset we use $\bi{x}_m = (h_m \theta_m,\phi_m)$, where we vary the magnitude field data $h_m$ from $0.25 J$ to $2 J$ using a regular step of $0.25 J$ (8 different values). In addition, we construct a regular rectangular ($50 \times 50$) grid for $(\theta_m, \phi_m)$. Then, our initial feature dataset encoded $2 \times 10^4$ different values for $\bi{x}_m$ ($m=1,2,...,2\times 10^4$). As before, we randomly select the $70$ \% of the dataset as the training data and the remaining $30\%$ for testing. In Table~\ref{table}, we show our results for the accuracy obtained for different numbers of qubits. During our simulations, we noted that DSK (GSK) always reaches accuracies over 98\% (91\%). These results confirm that GSK deteriorates when the number of qubits increases, as expected for a ground state method~\cite{anderson}. On the contrary, DSK always reaches the best accuracies and even outperforms the classical SVM algorithm in all cases. To summarize our findings, in Table~\ref{Table1}, we illustrate the main relevant aspects explored in this work regarding the two proposed methods (GSK and DSK).

\section{Conclusions}

We have proposed two strategies to calculate the Kernel function in the context of a dynamical critical problem in a multiqubit system with long-range interactions. First, when applying an external magnetic field, we characterized dynamical singularities using theoretical aspects related to symmetry recovery. We used the nonanalytic behavior of the rate function $\lambda(t)$ and the macroscopic magnetization $\langle M_x \rangle$ as an order parameter as predictors of dynamical singularities (binary classification). Then, after introducing the basic notion of data (features and labels) and classical Kernels, we extend these ideas to the quantum case of interest (closed and open systems). \par 

Making the analogy between a state vector of features (magnetic field orientation and amplitude) with the corresponding wave function, we introduced the concept of quantum Kernel. Since dynamical singularities of the rate function is a phenomenon that involves both critical properties of the ground state and dynamical aspects, we introduced both GSK and DSK methods as ideal candidates to solve the classification problem. Inspired by physical arguments and, testing our approach with a time-dependent Hamiltonian, Markovian bath, and analyzing the accuracy for an increasing number of qubits, we have demonstrated that a quantum Support Vector Machine (QSVM) algorithm that uses quantum states at a particular time is enough to reach accuracies over 98\% in different scenarios. These results outperform the classical SVM algorithm with automatic hyperparameter optimization implemented in MATLAB and show the advantage of using quantum machine learning algorithms to classify a property present in the same quantum states used by the algorithm. \par

The main novelty of this work is that it relies on dynamical prediction, unlike previous works that only focused on the ground state properties related to quantum phase transitions. We proved that our algorithm can learn how to classify a critical dynamical property by only knowing the system's ground state when the external field is present (GSK method). Moreover, the previous prediction can be improved if we use only one tomography of the state (DSK method). Further directions of this quantum algorithm and problem must include an in-depth analysis of the power-law scaling effect, the effectiveness against stochastic Hamiltonians, the role of the topology in the multiqubit system, and realistic implementation in a quantum computer.

\section{Acknowledgements}
D.T. acknowledges support from the Universidad Mayor through the Doctoral fellowship. A.N. acknowledges the financial support from the project Fondecyt Iniciaci\'on $\#$11220266. The authors acknowledge fruitful comments and observations from Francisco Albarrán-Arriagada.

\appendix

\section{Classic and Quantum Support Vector Classifier} \label{AppendixA}

\begin{figure}[ht!]
\centering
\includegraphics[width=1 \linewidth]{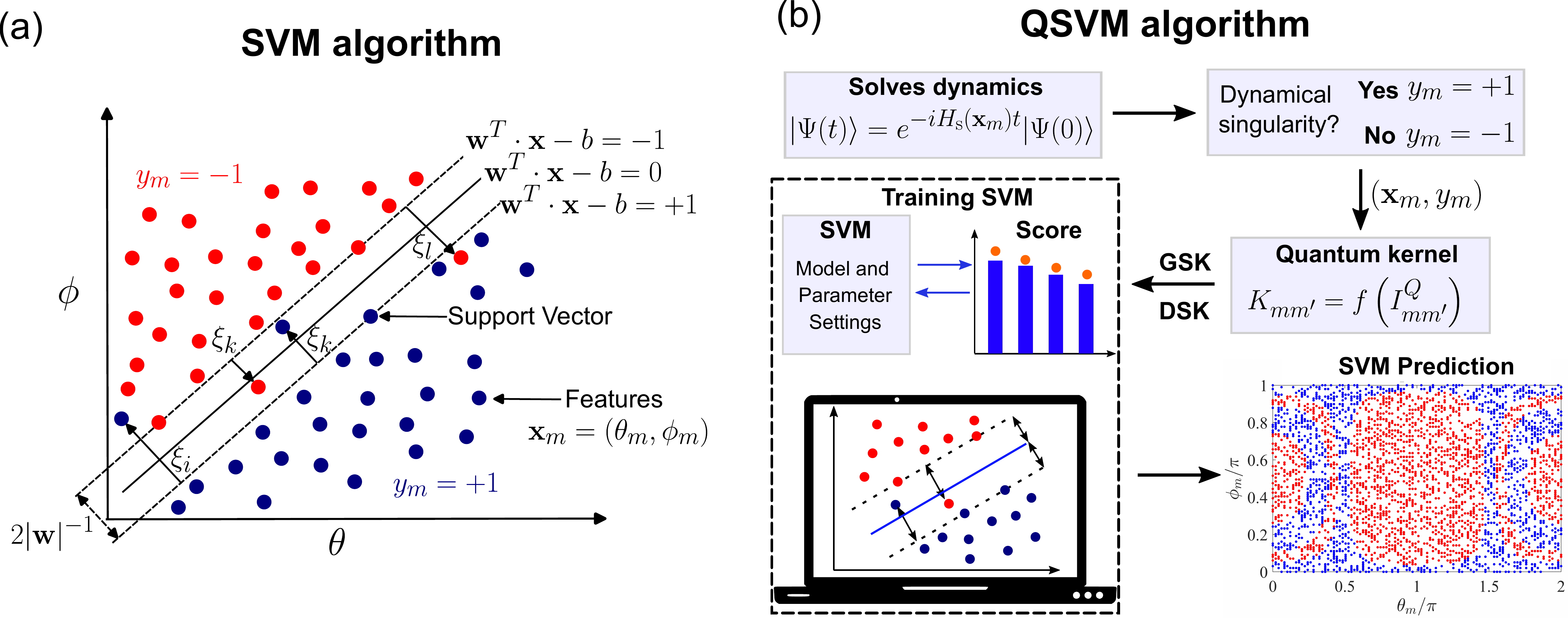}
\caption{(a) Schematic description of the SVC method, where $y_m$ is the label (two classes), $\bi{x}_m = (\theta_m , \phi_m)$ are the features, $\xi_m$ are the slack variables, $b$ is the bias, and $\bi{w}$ is the normal vector to the hyperplane. (b) QSVM method used in this work. The first step is estimating the system's state (wave function or density matrix) for a given set of features $\bi{x}_m$. The second and third steps correspond to the numerical calculations of the label and the quantum Kernel, respectively. Then, we use the quantum Kernels and dataset ($\bi{x}_m$ and $y_m$) to train our SVM algorithm to finally make predictions. } 
\label{SVM}
\end{figure}

In this appendix, we introduce the basic concepts of the classic and quantum Support Vector Machine (SVM) algorithm used in this work. First, it is essential to clarify that SVM is a supervised learning algorithm designed to solve classification (SVC for Support Vector Classification) and regression (SVR for Support Vector Regression) problems problems. Here, we only focus on SVC from both classical and quantum points of view. SVC aims to find the best hyperplane to separate data points optimally across different classes. Let us consider that we have a set of features given by the vectors $\bi{x}_m$ for a binary classification problem with labels $y_m = \pm1$ (two classes). As shown in figure~\ref{SVM}(a), ideally, the best hyperplane is the one that ensures a maximum distance from the nearest data points across all classes, implying a maximized margin ($2 \|\bi{w}\|^{-1}$). However, it is usual to introduce slack variables $\xi_m$ to improve the SVC method's performance when outlier points are present. Then, we move to the soft margin approach, where the optimal hyperplane is found by solving the optimization problem:

\begin{eqnarray} \label{SVC-optimization}
\min_{\bi{w},b,\xi} \left(\frac{1}{2} \|\bi{w}\|^2 + C \sum_{m=1}^{M} \xi_m \right), \quad 
\mbox{subject to}  \quad y_m(\bi{w}^T \bi{x}_m - b) \geq 1- \xi_m
\end{eqnarray}

where $\bi{w}$ is the normal vector to the hyperplane, $b$ is a number, and $C$ is the slack penalty (see figure.~\ref{SVM}(a)). Here, $b \|\bi{w}\|^{-1}$ gives the offset of the hyperplane from the origin along the normal vector $\bi{w}$. Also, the value of $C$ dictates the penalties imposed for misclassification of each data point, which is found using cross-validation. A diminished (higher) $C$ value implies lesser (higher) penalties, potentially leading to a broader (thin) decision margin but at the cost of increased (diminishing) misclassifications. \par

To solve the optimization problem~(\ref{SVC-optimization}) for nonlinear classification, one maps the initial data into a higher-dimensional space, enabling linear separation. To this end, it is expected to introduce Kernel functions $K(\bi{x}_m, \bi{x}_{m'})$ employing the Kernel trick. The kernel trick formula is given by

\begin{equation}
K(\bi{x}_m, \bi{x}_{m'}) = \varphi(\bi{x}_m) \cdot \varphi(\bi{x}_{m'})
\end{equation}

Here, $\varphi(\cdot)$ represents the transformation function of remapping input data to a higher-dimensional space. Notably, in the dual space, one can use the relations $\bi{w} = \sum_m \alpha_m y_m \varphi(\bi{x}_m)$ and $\bi{w} \cdot \varphi(\bi{x}) = \sum_m \alpha_m y_m k(\bi{x}_m,\bi{x})$, leading to the optimization problem in the dual space

\begin{eqnarray} \label{SVC-optimization-dual}
\max_{c_m} && \sum_m c_m - {1 \over 2} \sum_{m} \sum_{m'}y_m c_m(K_{mm'}) y_{m'}c_{m'}, \nonumber \\
\mbox{subject to} && \quad \sum_m c_m y_m =0\, \; \mbox{and} \; 0 \leq c_m \leq {1 \over 2 M \lambda}
\end{eqnarray}

where $\lambda$ is inversely related to $C$, $M$ is the amount of data ($m=1,...,M$), and $K_{mm'} = \varphi(\bi{x}_m) \cdot \varphi(\bi{x}_{m'})$ are the elements of the Gram matrix. In this new form, the optimization problem~(\ref{SVC-optimization-dual}) can be solved using quadratic computing with modern optimization techniques. \par

The quantum version of the algorithm is obtained by replacing the Kernel function with a proper metric between quantum states (wave function or density matrix). First, we solve the dynamics (Schr\"{o}dinger or master equations) for a given set of parameters $\bi{x}_m$ (magnetic field features). Second, by having the state as a function of time, we determine if, for a given feature $\bi{x}_m$, our system undergoes (or not) a dynamical singularity (by analyzing the probabilities in the ground state manifold). In simple words, we use the theoretical criteria explained in Sec.~\ref{Model} to determine the value of $y_m$ (binary classification problem). Then, we proceed to numerically calculate the quantum inner product $\bi{I}^{\rm Q}_{mm'}$ used for the quantum Kernel. In the closed case, we use the information given by the wave function. In particular, the GSK method uses the expression given in equation~(\ref{GSK}), while for the DSK method, we adopt equation~(\ref{GSK}). We use the information encapsulated in the density matrix for the open case. More precisely, for the GSK method we define the inner product as $\bi{I}^{\rm GSK}_{mm'}  = \mbox{Tr}(\rho_m(G) \rho_{m'}(G))$, where $\rho_m(G) = |G(\bi{h}_m)\rangle \langle G(\bi{h}_m) |$. Similarly, for the DSK, we use  $\bi{I}^{\rm GSK}_{mm'}  = \mbox{Tr}(\rho_m(t^c_m) \rho_{m'}(t^c_{m'}))$, where $\rho_m(t^c_m)$ is the density matrix at time $t^c_m$ such that the magnetization $\langle M_x \rangle  = \mbox{Tr}(M_x \rho)$ is maximized, as illustrated in figure~\ref{fig:Figure2}(b). Having calculated the Gram matrix for the GSK and DSK methods, we train our classic SVM algorithm by optimizing hyperparameters. Finally, using the best-trained SVC model, we conclude with predictions using a testing dataset not considered in the initial training step. A detailed description of the QSVM method is given in figure~\ref{SVM}(b).

\section{Digital quantum simulation of the proposed kernels} \label{AppendixB}
The time evolution operator used in this work is defined as $U(t) = \mbox{exp}(-i H_{\rm s}(\bi{h}))$, where $H_{\rm s}(\bi{h})$ is the Hamiltonian given in equation~(\ref{Hamiltonian}). The following equation gives a possible decomposition of the operator $U(t_m^{\rm c})$ using gates to first order in time (trotterization)

\begin{eqnarray} 
    U(\bi{h}_m,t_m^c) &\approx& \prod_{k=1}^n\prod_{i\neq j}R_{XX}^{i,j}(-2J_{ij}t_m^c/n) \nonumber \\
    && \times 
    \prod_{i=1}^{N}R_X^{i}(\Delta_{x,mn})R_Y^{i}(\Delta_{y,mn})R_Z^{i}(\Delta_{z,mn}), 
\end{eqnarray}

where $\Delta_{\beta,mn} = -2h_{\beta} t_m^c/n$ ($\beta = x,y,z$) and we decomposed the time in $n$ steps. Here, $R_\beta$ and $R_{\beta \beta}$ are rotation gates with Pauli matrix operator in the following form

\begin{eqnarray}
    R_{\sigma}(\theta) = \exp\left(-\frac{\theta}{2}\sigma\right),\ R_{\sigma\sigma}(\theta)= \exp\left(-\frac{\theta}{2}\sigma \otimes \sigma\right),
\end{eqnarray}

with $\sigma = X, Y, Z$ are the Pauli gates. Then, the dynamical state Kernel (DSK) can be obtained using the inversion test by calculating the probability of measure $\ket{0}^{\otimes N}$ in the following state

 \begin{eqnarray}
     \ket{\Psi(m,m')} = U^{\dag}(\bi{h}_m,t_m^c)U(\bi{h}_{m'},t_{m'}^c)\ket{0}^{\otimes N}.
 \end{eqnarray}

\section{Effect of the phase damping channel} \label{AppendixC}

\begin{figure}[ht!]
\centering
\includegraphics[width=1.05 \linewidth]{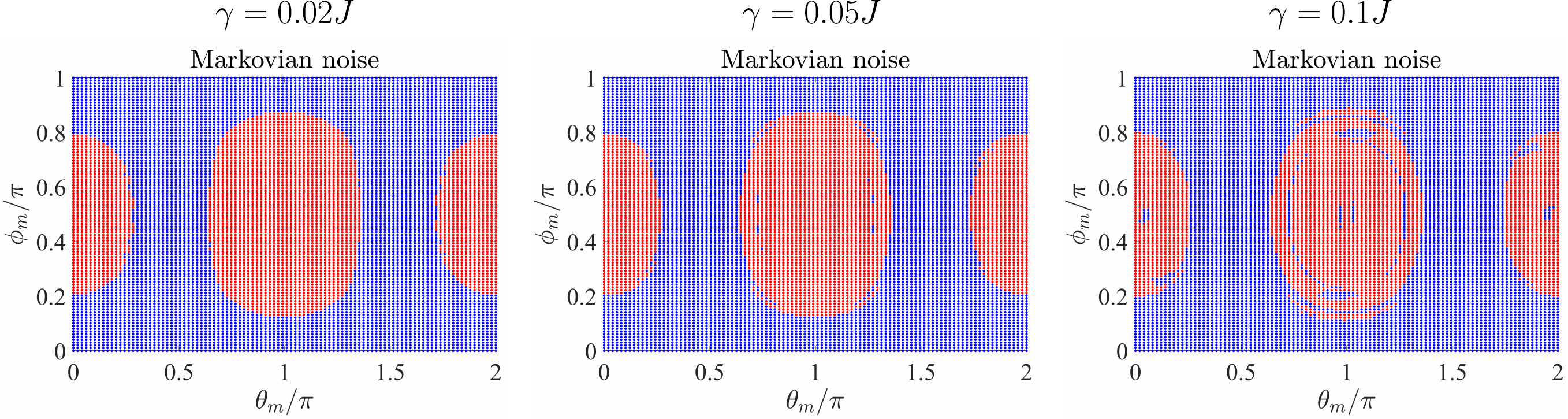}
\caption{Angular distribution of the binary classification problem under the effect of phase damping channel with different dissipative rates.} 
\label{Figure5}
\end{figure}

Using the Markovian master equation given in equation~(\ref{ME}) and interpreting $L_k = \sigma_z^{k} = \ket{\uparrow}_k\bra{\uparrow}-\ket{\downarrow}_k\bra{\downarrow}$, we can model a phase damping channel with a rate $\gamma_k = \gamma$. In figure~\ref{Figure5}, we observe the dynamical singularities pattern for three different values of the phase damping rate. We note that a phase damping channel generates an interesting effect. Based on the pattern distribution, we note that in the regime $\gamma < 0.1J$, it is expected to have a good performance of our proposed DKS method since angular distributions do not reveal a more complex pattern than others presented in the main text. \an{For example, we use our algorithm for the case $\gamma = 0.02J$, obtaining an accuracy equal to $97.6\%$ using the DSK, which demonstrates the adaptability of our method for different dissipative channels.}

\section*{References}

\bibliographystyle{unsrt}

\begin{thebibliography}{9}
\bibitem{SM1}
In the appendix, we explain details associated with the classical and quantum implementations of Support Vector Machines for classification problems.
\end{thebibliography}

\end{document}